\documentclass[12pt]{article}
\usepackage{amsfonts}
\usepackage{amssymb}
\usepackage{graphics,psboxit,amsmath}

\def\hybrid{\topmargin -20pt    \oddsidemargin 0pt
        \headheight 0pt \headsep 0pt
        \textwidth 6.35in       
        \textheight 9.25in       
        \marginparwidth .875in
        \parskip 5pt plus 1pt   \jot = 1.5ex}

\hybrid

\def\baselinestretch{1.2}

\catcode`\@=11

\def\marginnote#1{}
\newcount\hour
\newcount\minute
\newtoks\amorpm
\hour=\time\divide\hour by60
\minute=\time{\multiply\hour by60 \global\advance\minute by-\hour}
\edef\standardtime{{\ifnum\hour<12 \global\amorpm={am}%
        \else\global\amorpm={pm}\advance\hour by-12 \fi
        \ifnum\hour=0 \hour=12 \fi
        \number\hour:\ifnum\minute<10 0\fi\number\minute\the\amorpm}}
\edef\militarytime{\number\hour:\ifnum\minute<10 0\fi\number\minute}

\def\draftlabel#1{{\@bsphack\if@filesw {\let\thepage\relax
   \xdef\@gtempa{\write\@auxout{\string
      \newlabel{#1}{{\@currentlabel}{\thepage}}}}}\@gtempa
   \if@nobreak \ifvmode\nobreak\fi\fi\fi\@esphack}
        \gdef\@eqnlabel{#1}}
\def\@eqnlabel{}
\def\@vacuum{}
\def\draftmarginnote#1{\marginpar{\raggedright\scriptsize\tt#1}}

\def\draft{\oddsidemargin -.5truein
        \def\@oddfoot{\sl preliminary draft \hfil
        \rm\thepage\hfil\sl\today\quad\militarytime}
        \let\@evenfoot\@oddfoot \overfullrule 3pt
        \let\label=\draftlabel
        \let\marginnote=\draftmarginnote
   \def\@eqnnum{(\theequation)\rlap{\kern\marginparsep\tt\@eqnlabel}%
\global\let\@eqnlabel\@vacuum}  }

\def\preprint{\twocolumn\sloppy\flushbottom\parindent 2em
        \leftmargini 2em\leftmarginv .5em\leftmarginvi .5em
        \oddsidemargin -.5in    \evensidemargin -.5in
        \columnsep .4in \footheight 0pt
        \textwidth 10.in        \topmargin  -.4in
        \headheight 12pt \topskip .4in
        \textheight 6.9in \footskip 0pt
        \def\@oddhead{\thepage\hfil\addtocounter{page}{1}\thepage}
        \let\@evenhead\@oddhead \def\@oddfoot{} \def\@evenfoot{} }

\def\numberbysection{\@addtoreset{equation}{section}
        \def\theequation{\thesection.\arabic{equation}}}

\def\underline#1{\relax\ifmmode\@@underline#1\else
        $\@@underline{\hbox{#1}}$\relax\fi}

\def\titlepage{\@restonecolfalse\if@twocolumn\@restonecoltrue\onecolumn
     \else \newpage \fi \thispagestyle{empty}\c@page\z@
        \def\thefootnote{\fnsymbol{footnote}} }

\def\endtitlepage{\if@restonecol\twocolumn \else \newpage \fi
        \def\thefootnote{\arabic{footnote}}
        \setcounter{footnote}{0}}  

\catcode`@=12
\relax

\def\figcap{\section*{Figure Captions\markboth
        {FIGURECAPTIONS}{FIGURECAPTIONS}}\list
        {Figure \arabic{enumi}:\hfill}{\settowidth\labelwidth{Figure
999:}
        \leftmargin\labelwidth
        \advance\leftmargin\labelsep\usecounter{enumi}}}
 \relax
\def\tablecap{\section*{Table Captions\markboth
        {TABLECAPTIONS}{TABLECAPTIONS}}\list
        {Table \arabic{enumi}:\hfill}{\settowidth\labelwidth{Table
999:}
        \leftmargin\labelwidth
        \advance\leftmargin\labelsep\usecounter{enumi}}}
 \relax
\def\reflist{\section*{References\markboth
        {REFLIST}{REFLIST}}\list
        {[\arabic{enumi}]\hfill}{\settowidth\labelwidth{[999]}
        \leftmargin\labelwidth
        \advance\leftmargin\labelsep\usecounter{enumi}}}
 \relax
\makeatletter
\newcounter{pubctr}
\def\publist{\@ifnextchar[{\@publist}{\@@publist}}
\def\@publist[#1]{\list
        {[\arabic{pubctr}]\hfill}{\settowidth\labelwidth{[999]}
        \leftmargin\labelwidth
        \advance\leftmargin\labelsep
        \@nmbrlisttrue\def\@listctr{pubctr}
        \setcounter{pubctr}{#1}\addtocounter{pubctr}{-1}}}
\def\@@publist{\list
        {[\arabic{pubctr}]\hfill}{\settowidth\labelwidth{[999]}
        \leftmargin\labelwidth
        \advance\leftmargin\labelsep
        \@nmbrlisttrue\def\@listctr{pubctr}}}
 \relax
\makeatother
\newskip\humongous \humongous=0pt plus 1000pt minus 1000pt

\newif\ifdtup

\relax

\def\be{\begin{equation}}
\def\ee{\end{equation}}
\def\ba{\begin{eqnarray}}
\def\ea{\end{eqnarray}}

\def\no{\noindent}

\def\IR{\relax{\rm I\kern-.18em R}}
\def\II{\relax{\rm 1\kern-.35em1}}

\renewcommand{\theequation}{\thesection.\arabic{equation}}
\csname @addtoreset\endcsname{equation}{section}

\def\IR{\relax{\rm I\kern-.18em R}}
\def\inv{^{\raise.15ex\hbox{${\scriptscriptstyle -}$}\kern-.05em 1}}

\begin{document}

\begin{titlepage}
\begin{center}

\hfill IFT-UAM/CSIC-09-06\\
\vskip -.1 cm
\hfill arXiv:0902.1427\\

\vskip .5in

{\LARGE Wrapping in maximally supersymmetric and marginally deformed 
${\cal N}=4$ Yang-Mills}
\vskip 0.4in

{\bf Johan Gunnesson}
\vskip 0.1in

Instituto de F\'{\i}sica Te\'orica UAM-CSIC, C-XVI \\
Universidad Aut\'onoma de Madrid,
Cantoblanco, 28049 Madrid, Spain\\
{\footnotesize{johan.gunnesson@uam.es}}

\end{center}

\vskip .4in

\centerline{\bf Abstract}
\vskip .1in
\no
In this note we give evidence for an equality of the spectra, 
including wrapping, of the $SU(2)$-sector spin chain for real 
deformations $\beta$ and $\beta+1/L$, in marginally $\beta $-deformed ${\cal N}=4$ Yang-Mills, 
which appears after relaxing the cyclicity constraint. Evidence 
for the equality is given by evaluating the first 
wrapping correction to the energy of the undeformed magnon of momentum 
$\pi$, and the $\beta=1/2$, physical magnon, for several spin chain 
lengths $L$. We also show that the term of maximal transcendentality 
coincides for both magnons to all $L$. As a by-product we provide an 
expression for the first wrapping correction to the $\beta = 1/2$ single-magnon 
operator dimension, valid for all even $L$. We then apply the 
symmetry to the magnon dispersion relation of ${\cal N}=4$, obtaining 
its first wrapping correction for a discrete set of magnon momenta.

\vskip .4in
\noindent

\end{titlepage}
\vfill
\eject

\def\baselinestretch{1.2}


\baselineskip 20pt


\section{Introduction}

Using the techniques of integrability, much progress has been made towards confirming the complete identification of the planar spectra of $\mathcal{N}=4$ gauge theory and $AdS_5 \times S^5$ string theory implied by the AdS/CFT correspondence \cite{AdSCFT}. First, the construction of the scattering matrix and the asymptotic Bethe ansatz \cite{BDS,ABA} turned reachable the spectrum of asymptotically long operators and infinitely 
fast strings in the supergravity background. For short operators, or strings with small charges it was shown that the Bethe ansatz equations are no longer valid (see for example \cite{LipatovStaudacher}). In the gauge theory wrapping interactions need to be taken into account when the perturbative order reaches the length $L$ of the gauge invariant operator \cite{SiegWrapping} while in the string theory finite-size effects come from virtual corrections traveling around the worldsheet cylinder \cite{AmbjornJanik}. Recently, however, even wrapping effects have started to become manageable. The four-loop anomalous dimension of the Konishi operator, which is the simplest case where wrapping effects are present, has been obtained
both from perturbative computations on the gauge theory side \cite{gaugeKonishi} and from a thermodynamic Bethe ansatz calculation based on
the quantum field theory of the string worldsheet \cite{JanikKonishi}. This last approach was subsequently extended to the entire spectrum of twist-two operators \cite{Janik}. Finally, there is a recent proposal, also based on integrable two-dimensional QFT, for a set of equations determining the wrapping corrections to the entire spectrum of planar, gauge invariant operators \cite{GromovKazakov}.

Given this success in the maximally supersymmetric case, one would like to find other examples of gauge/gravity correspondences which exhibit integrability. One possibility is to deform the $\mathcal{N}=4$ theory in such a way that the maximum amount of integrability is preserved. With this in mind, one can consider marginally $\beta$-deformed $\mathcal{N}=4$ theory \cite{LeighStrassler}, which has integrable subsectors \cite{BerensteinCherkis,Beiserttwist}, admitting an asymptotic Bethe ansatz. The $\beta$-deformed theory is obtained from the $\mathcal{N}=4$ theory by modifiying the superpotential as
\be
\text{Tr}([X,\, Y],\, Z) \rightarrow e^{i\pi \beta}\text{Tr}(X,\, Y,\, Z) - e^{-i\pi \beta}\text{Tr}(Y,\, X,\, Z) \ .
\ee
The deformation $\beta$ can be complex in general, but we will limit ourselves to the study of real $\beta$, for which the theory remains superconformal \cite{deformeddispersion} and probably completely integrable. The string theory dual of this theory was constructed in \cite{LuninMaldacena}, and is obtained by deforming the $S^5$ part of the background through a series of T-duality transformations and shifts of angular variables. The integrable structure of the string theory was analyzed in \cite{Frolov}. Wrapping calculations have been performed in the deformed theory for operators corresponding to one and two magnons \cite{deformedwrapping, Zanon}, and the first finite size correction to the giant magnon dispersion relation has been obtained on the string side \cite{BykovFrolov}.

In this note we will see that at least in the $SU(2)_\beta$-sector, consisting of scalar operators of the form
\be
\text{Tr}(\Phi Z Z \Phi \cdots ) \ ,
\ee
the energies, including their wrapping corrections, have a certain symmetry, which relates them at different values of the deformation. One can then use this symmetry to obtain additional information on the original $\mathcal{N}=4$ theory. In particular, we will extract the first wrapping correction to the magnon dispersion relation (for a discrete set of momenta), which can give insights into the structure of the wrapping dependence of the dilatation operator.

To be more precise, we will provide evidence that for a spin chain of length $L$ 
deformations differing by $1/L$ are equivalent. 
The equality of spectra at $\beta$ and $\beta + 1/L$ 
is obvious in the twisted asymptotic Bethe ansatz, and not difficult 
to see in the asymptotic Hamiltonian. It is however not clear whether wrapping 
respects the equivalence, but we will show that 
the first contribution to wrapping in certain cases is consistent with it.

The equality is, however, not directly visible from the gauge theory 
since spin chain states that satisfy the cyclicity constraint, which is the condition that physical, single trace operators must satisfy, 
at deformation $\beta$ will generally not do so at $\beta + 1/L$. 
But the equality of spectra is still useful and we will see, as mentioned above, how it can be employed to 
extract information on the magnon dispersion relation and the physical 
spectrum. One can also combine it with the symmetry $\beta \rightarrow -\beta$ 
in order to relate the spectra at several inequivalent values of the deformation.

In section two we will show why the spin chain spectra is the 
same at $\beta$ and $\beta + 1/L$ in the asymptotic case. We also mention 
a wrapping correction that satisfies the symmetry and note how 
the string theory finite size corrections are consistent with it. In 
section three we provide further evidence that wrapping respects the symmetry, 
showing the equality of the energies of 
the $L=4,\, 6,\, 8$ magnon at $\beta = 0$, $p=\pi$ and $\beta = 1/2$, $p = 0$, 
and that the coefficient of the maximal trascendentality term is the same for all $L$. As an application, 
we note that the equality of these energies provides an efficient 
way of calculating the first wrapping correction to the physical 
magnon operator anomalous dimension at $\beta = 1/2$. We end the section by using the symmetry to obtain the magnon dispersion relation of $\mathcal{N}=4$ for some specific values of the momentum. We then present some concluding remarks in section four.


\section{Equality of the spectra at $\beta $ and $\beta + 1/L$}

\no
In this section we will show that the spectra for the $SU(2)_\beta$ 
spin chain at deformation $\beta$ is the same as that for deformation 
$\beta +1/L$. We will begin by looking 
at the Bethe ansatz, after which we will study the structure of the hamiltonian. 
Let us for convenience consider in detail the one-loop ansatz \cite{BerensteinCherkis}. 
The discussion carries over unaltered to the asymptotic 
all-loop case, since the deformation enters in the same way 
\cite{FrolovRoibanTseytlin,Beiserttwist}. The marginally deformed 
one-loop $SU(2)$ Bethe ansatz, with real 
deformation parameter $\beta$, is given by
\be
e^{-2\pi i \beta L}\left( \frac{u_k + i/2}{u_k - i/2} \right)^L = \prod _{j \neq k =1}^M 
\frac{u_k - u_j + i}{u_k - u_j -i} \ , \label{eq:BA}
\ee
with cyclicity contraint
\be
\prod _{k=1}^M \frac{u_k + i/2}{u_k - i/2} = e^{2\pi i \beta M} \ , \label{eq:cyclicity}
\ee
and magnon energy
\be
E_k = \frac{2g^2}{u_k^2 + 1/4} \ . \label{eq:ek}
\ee
When $\beta \rightarrow \beta + 1/L$, the deformation factor 
$e^{-2\pi i \beta L}$ is unaltered in \eqref{eq:BA}. Thus, any 
set of rapidities solving the Bethe equations for deformation 
$\beta$ will also do so for $\beta + 1/L$, and vice versa. Furthermore, 
\eqref{eq:ek} shows that the solution has the same energy in both cases. 
The only thing that can be violated is the cyclicity constraint 
\eqref{eq:cyclicity}, which implies that the total momentum of 
the state is shifted \footnote{This assumes that we are using the 
picture of a periodic spin chain, so that physical operators 
correspond to zero total momentum. Alternatively, one can eliminate 
the deformation dependence of the dispersion relation and the S-matrix 
by introducing twisted boundary conditions \cite{BerensteinCherkis,FrolovRoibanTseytlin}. 
In the latter case, the momentum is not altered upon changing $\beta$, and the 
cyclicity constraint instead gives a $\beta$-dependent momentum condition for 
physical operators.} by $-2\pi M/L$ on going from $\beta$ to 
$\beta + 1/L$. Physical states (satisfying the cyclicity constraint) 
at deformation $\beta$ will therefore correspond to unphysical states at 
deformation $\beta + 1/L$. We thus find a symmetry of the more general spin chain, 
with the cyclicity constraint removed, which the gauge theory is embedded in. 
In what follows we will be particularly interested in the symmetry 
$\beta \rightarrow \beta + 1/2$, present for even length spin chains 
(obtained by applying the general symmetry $L/2$ times). The cyclicity 
constraint then implies that physical states with odd magnon number in 
one case will correspond to unphysical states of total momentum $p=\pi$ in the other.  

Let us now exhibit the symmetry between the different values of $\beta$ at the level of 
the spin chain Hamiltonian. The undeformed 
Hamiltonian can be written as a linear combination of generalized 
permutations \cite{BCS},
\be
\{ a_1, \ldots , \, a_n \} \equiv \sum _{l=0}^{L-1}P_{a_1 +l,\, 
a_1 + l +1}\cdots P_{a_n +l,\, a_n + l +1} \ ,
\ee
where the $P_{i,\, i+1}$ permutes the spins at sites $i$ and $i+1$ 
and can be written in terms of the Pauli matrices as
\be
P_{i,\, i+1} = \frac{1}{2}\left( \II_{i,\, i+1} + \sigma ^z_i 
\otimes \sigma ^z_{i+1} + \sigma ^+_i \otimes \sigma ^-_{i+1} + \sigma ^-_i 
\otimes \sigma ^+_{i+1}  \right) \ . \label{eq:perm}
\ee
The asymptotic deformed Hamiltonian can be obtained from the undeformed one 
by simply replacing all permutations \eqref{eq:perm} with deformed 
operators \cite{BerensteinCherkis, deformedwrapping, FrolovRoibanTseytlin},
\be
\boldsymbol{P}_{i,\, i+1} = \frac{1}{2}\left( \II_{i,\, i+1} + \sigma ^z_i 
\otimes \sigma ^z_{i+1} + e^{2\pi i \beta}\sigma ^+_i \otimes \sigma ^-_{i+1} 
+ e^{-2\pi i \beta}\sigma ^-_i \otimes \sigma ^+_{i+1}  \right) \ . \label{eq:permdef}
\ee
The coefficients that multiply the generalized permutations do not change 
when going to the deformed theory. Now, making the change 
$\beta \rightarrow \beta + 1/L$ produces additional phases multiplying 
the last two terms in \eqref{eq:permdef}, but which can be removed by 
performing a non-local change of basis, such as the one in \cite{BerensteinCherkis}. 
Normally, such a change of basis introduces an additional twist in the boundary 
conditions, but when the change in deformation is $1/L$, the twist becomes unity.
This can also be combined with the more obvious symmetry 
\be
\beta \rightarrow -\beta\; , \;\;\; p_k \rightarrow -p_k \, ,
\ee 
which can be read directly from \eqref{eq:permdef}, since the reversal of 
$\beta$ can be undone by changing the order of the sites, or equivalently, 
by changing the sign of the magnon momenta. For $\beta = 0$ and $\beta = 1/2$ 
this reduces to the ordinary parity symmetry. 

It is however not clear whether wrapping effects will 
spoil the correspondence between $\beta$ and $\beta + 1/L$. One would 
therefore like some wrapping computations respecting the 
symmetry. As a first example, the dimensions of all even length operators 
with even number of impurities $M$ should coincide at $\beta=0$ 
and $\beta=1/2$, since the cyclicity constraint is not violated in that 
case. In particular, the $M=2$ operator should respect the symmetry. 
In \cite{deformedwrapping} the anomalous dimension of the two-impurity operator is given 
as a function of $\beta$, for $L=4$, and the deformation enters through 
$\cos (4\pi \beta)$, invariant under both 
$\beta \rightarrow \beta + 1/2$ and $\beta \rightarrow -\beta$. In 
the next section we will also show the equality of two single-magnon 
energies at $\beta = 0$ and $\beta = 1/2$. 

Further evidence in favor of the symmetry arises from string theory. The first 
finite size correction to the magnon dispersion relation in the 
$\beta$-deformed theory was calculated in \cite{BykovFrolov}. At strong-coupling 
the deformation enters through a quantity $\cos \Phi$, where
\be
\Phi = \frac{2\pi (n_2 - \beta L)}{2^{3/2}\cos ^3 \frac{\tilde{p}}{4}} \ , \label{eq:Phi}
\ee
where $n_2$ is a winding number, which in order for the existence of a solution to the equations of motion, 
had to be chosen as $[ \beta L ]$, the integer nearest to $\beta L$. The quantity $\tilde{p}$ is the momentum 
when twisted boundary conditions are used, or if one prefers periodic boundary conditions is related to the periodic momentum $p$ via 
\be
\tilde{p} = p + 2\pi \beta \ . \label{eq:ptilde}
\ee
The momentum $\tilde{p}$ is invariant under the symmetry, while the complete phase 
\eqref{eq:Phi} is so (in addition to the $\beta \rightarrow -\beta$ symmetry) when the winding is set to $[ \beta L ]$. Different choices of the winding, such as a fixed number, would violate the symmetry for general $\tilde{p}$.


\section{Equality of magnon energies}

\label{sec:magnon}

\no
An essential ingredient in the asymptotic Bethe ansatz of marginally 
$\beta$-deformed $\mathcal{N}=4$ Yang-Mills is the magnon dispersion relation 
\cite{BDS,FrolovRoibanTseytlin},
\be
E = -1 + \sqrt{1 + 16g^2 \sin ^2 \left( \frac{\tilde{p}}{2} \right)} \ , 
\label{eq:dispersion}
\ee
where $g^2 \equiv g^2_{\text{YM}}N/16\pi^2$ and $\tilde{p}$ is defined in the previous section.
For a non-zero $\beta$-deformation, single-magnon operators
\be
\text{Tr} \left( \Phi Z^{L-1} \right) 
\label{eq:op1magnon}
\ee
are not protected by supersymmetry and thus acquire non-zero anomalous dimension
given by \eqref{eq:dispersion} with $\tilde{p} = 2 \pi \beta$. 
As \eqref{eq:dispersion} does not explicitly depend on $\beta$ it is invariant 
under the symmetry $\beta \rightarrow \beta + 1/L$. 
In this section we will show the equality of the first wrapping correction to the 
energies of the physical $\beta = 1/2$ magnon, and the $\beta = 0$, $p = \pi$ magnon, 
for even $L$, as expected from the symmetry. We will first study the case $L=4$ 
and then move on to larger $L$, relying on the formula 
recently presented in \cite{Zanon} for 
the first wrapping contribution to the anomalous dimensions of the single magnon 
operators \eqref{eq:op1magnon} in the $\beta$-deformed theory. 
It was shown that the first wrapping correction, appearing at order $g^{2L}$, takes 
the general form 
\be
\delta \gamma _L ^\beta = -2L g^{2L}\Big[ \big( C_{L,0}(\beta) 
- C_{L,L-1}(\beta) \big) P_L - 2\sum_{j=0}^{\left\lfloor \frac{L}{2}\right\rfloor - 1} 
\big(C_{L,j}(\beta) - C_{L,L-j-1}(\beta) \big) I_L^{(j+1)}  \Big] \ ,  \label{eq:wrapZanon}
\ee
where
\be
C_{L,j}(\beta) = -8 \sin ^2(\pi \beta) \cos \left[ 2\pi \beta (L - j -1) \right] \ , \quad
P_L = \frac{2}{L}\binom{2L-3}{L-1} \zeta (2L-3) \ . \label{eq:C}
\ee
The integrals $I_L^{(j+1)}$ are presented explicitly up to nine-loops in \cite{Zanon}. 
When $\beta = 1/2$ one has
\be
\delta \gamma _L ^{1/2} = -16L g^{2L}\left(1 - (-1)^{(L-1)}\right) 
\Big[ P_L - 2\sum_{j=0}^{\left\lfloor\frac{L}{2}\right\rfloor 
- 1} (-1)^j I_L^{(j+1)}  \Big] \ .  \label{eq:wrapZanon1/2}
\ee
Interestingly, the wrapping contribution vanishes for all odd operator 
lengths when $\beta = 1/2$. Furthermore, using (see the appendix in \cite{Zanon})
\be
I_4^{(1)} = \frac{1}{2}\zeta (3) + \frac{5}{2}\zeta (5) \ , \quad 
I_4^{(2)} = -\frac{3}{2} \zeta (3) + \frac{5}{2} \zeta (5) \ ,
\ee 
we get the first non-trivial wrapping correction to the operator of length four as
\be
\delta \gamma_4^{1/2} = 512 \zeta(3) -640 \zeta(5) \ .\label{eq:zanonbeta1/2r4}
\ee

In the $\beta$-deformed theory, one can thus obtain some non-trivial 
information about wrapping corrections to the magnon dispersion 
relation solely from the physical spectrum. By constrast, this would 
seem impossible in the original $\mathcal{N}=4$ theory. 
However, in \cite{GGH} it was suggested that this not be the case. 
Instead it appears to be possible to extract 
the energy of a magnon of momentum $p=\pi$ from the physical spectrum of twist-two operators. 
To be more precise, the anomalous dimension of twist-two operators
\be
\text{Tr}\left(Z D^M Z \right) \ ,
\ee
can be expanded perturbatively as
\be
\gamma (M) = \sum_{r=1}^{\infty} \gamma_{r}(M) g^{2r} \ ,
\ee
where the $\gamma_r (M)$ can be conveniently expressed through a basis of harmonic sums, defined through
\be
S_a \equiv \sum_{j=1}^M \frac {(\hbox{sgn}(a))^j}{j^a} \ , \quad 
S_{a_1,\ldots,a_n} \equiv \sum_{j=1}^M \frac {(\hbox{sgn}(a_1))^j}{j^{a_1}} 
S_{a_2,\ldots,a_n}(j) \label{eq:harm} \ .
\ee
Setting $M=1$ into the three-loop expressions\cite{KLV} one finds \cite{GGH}, 
\be
\gamma(1) = 8 g^2 - 32 g^4 + 256 g^6 + \cdots ,
\ee
which coincides with the expansion of the dispersion relation 
\eqref{eq:dispersion} at $p=\pi$. Furthermore 
setting $M=1$ into the four-loop term in the asymptotic Bethe ansatz formula \cite{LipatovStaudacher} 
this trend continues. We can thus interpret $\gamma (1)$ as giving the energy 
of a single magnon of momentum $p=\pi$.\footnote{Indeed, at $M=1$, the 
one-loop Baxter equation is given by
\be
\left(u+i/2 \right)^L P_M (u+i)+\left( u-i/2 \right)^L P_M(u-i) = t(u)P_M(u) \ ,
\ee
where 
\be
P_M(u) \propto \prod_{i=1}^M (u-u_i) \ ,
\ee
given in terms of the Bethe roots $\{ u_i \}$, and the transfer matrix
\be
t(u)= 2u^L + \cdots \ ,
\ee
is a polynomial where the omitted terms, depending on $M$ and $L$, are of 
lower powers in $u$. The only non-trivial solution for $M=1$ and $L$ even, which follows 
from requiring that the highest powers of $t(u)$ be $2u^L$, is that $P_1(u)\propto u$. 
We thus have a single $u=0$ root at even $L$, corresponding to $p=\pi$.} 
Therefore if it continues to hold to higher orders that $\gamma(1)$ is the energy of a single 
magnon with momentum $p = \pi$ then 
we should be able to extract non-trivial information about the magnon dispersion relation directly 
from the physical spectrum. The way this bypasses the physical zero-momentum constraint is that 
only even values of $M$ correspond to physical, gauge theory states. 
Odd $M$ are unphysical, as is the $p=\pi$ state found at $M=1$. Information 
about such unphysical states can be obtained by analytically continuing 
from the physical values of $M$. In fact the prescription used for analytically continuing 
the harmonic sums is the $(-)$ prescription 
of \cite{KotikovVelizhanin}.
and coincides with \eqref{eq:harm} at odd values of $M$. 
By contrast, usually the physical $(+)$ prescription is used in 
$\mathcal{N}=4$ Yang-Mills, since it connects the physical, even $M$ states.

Starting from four loops wrapping contributions will affect the 
twist-two anomalous dimension. It was 
suggested in \cite{GGH} that such contributions would cancel 
when $M=1$, and therefore would not affect the dispersion relation for 
$p = \pi$. However, the complete spectrum of twist-two operators 
at four-loops including wrapping effects has 
been computed recently \cite{Janik}. The wrapping contributions are given by
\begin{align}
\delta \gamma _{4}(M) &= 256 S_1^2 (-S_5 + S_{-5} + 2 S_{4, 1} - 2 S_{3, -2} + 
        2 S_{-2, -3} - 4 S_{-2, -2, 1}) - \nonumber \\
  &-512 S_1^2 S_{-2} \zeta(3) - 640 S_1^2 \zeta(5) \ .
\end{align}
In the single magnon case this expression becomes \footnote{The non-vanishing irrational contribution was not expected in \cite{GGH}.}
\be
\gamma _{4,\text{wrapping}}(1) = 512 \zeta(3) -640 \zeta(5) \ , \label{eq:janik4loopm1}
\ee
which coincides with the wrapping contribution \eqref{eq:zanonbeta1/2r4} to the magnon operator at $\beta = 1/2$.

We have thus verified that the $p=\pi$ magnon in $\mathcal{N}=4$ Yang-Mills and the $\beta = 1/2$ physical magnon 
have the same energy up to four-loops, including wrapping effects, 
with the length of the spin chain such that wrapping starts at four loops. 
This is consistent with the $\beta \rightarrow \beta + 1/2$ symmetry. 

We will shortly verify also the cases $L=6$ and $L=8$. 
But let us first simply note that an important issue in evaluating the 
wrapping contribution of \cite{Janik} is that special 
care must be taken at the odd values of $M$. To make sure that wrapping starts at the correct 
perturbative order an additional $(-1)^M$ factor must be introduced at a certain 
point in the expression for the wrapping contribution. 
This prescription is a departure from the physical analytical continuation, 
and seems to be equivalent to a re-insertion of the $(-1)^M$ factor in the definition of the alternating 
harmonic sums. The main consequence of this is that the wrapping 
contributions obtained in \cite{Janik} at odd values of $M$ correspond 
to the $(-)$ analytical continuation, which is precisely the prescription to be applied in our case.

%
In what follows we will extend the previous results to longer spin chains. 
The formula \eqref{eq:wrapZanon1/2} can be applied to extract the first 
wrapping correction to the $\beta = 1/2$ magnon energy for arbitrarily large 
$L$, but since the integrals $I_L^{(j+1)}$ are given explicitly up to $L=9$ 
in \cite{Zanon} we will mainly limit ourselves to studying the first 
cases. We will however check that the transcendentality structures 
of the two magnon energies coincide also to higher $L$, and show that 
the coefficients of the maximal transcendentality terms match to all $L$. 
For the $p=\pi$, undeformed case, one can obtain the wrapping correction 
for longer spin chains from the equations of \cite{Janik} by simply 
chosing different values for $L$ in the exponential term. In fact, as 
shown in appendix \ref{app:Janik}, for the single magnon, and $L$ even, 
the formula for the wrapping correction takes the simple form\footnote{A closed form for this expression, which answers many of the questions raised below, was produced in \cite{Beccariamagnon}.}
\be
\delta E(L) \! = \! -256 \cdot 4^{L-2} i g^{2L} \! \! 
\sum _{Q=1}^{\infty} 
\text{Res}_{q=iQ}\left[\frac{Q^2 \left(1 - 1/(q^2 + Q^2)\right)^2}{ 
(q^2 \! + \! (Q - 1)^2) (q^2 \! + \! (Q \! + \! 1)^2) (q^2 \! + \! 
Q^2)^{L-2} } \right] \ \label{eq:wraphigherL} ,
\ee
where we have chosen to write this in terms of the $SU(2)$-sector 
length, explaining why it enters as $L-2$. Evaluating this expression for $L=6$ 
and $L=8$ one has
\ba
\delta E(6) & = & 128 \big[ 32 \zeta (5) + 28 \zeta (7) - 63 \zeta (9) \big] \ , \\
\delta E(8) & = & 
768 \big[ 32 \zeta(7) + 64 \zeta(9) + 44 \zeta(11) - 143 \zeta(13) \big] \ .
\ea
These expressions coincide with what is obtained from \eqref{eq:wrapZanon1/2}.

For higher $L$ we can also check that \eqref{eq:wrapZanon} and 
\eqref{eq:wraphigherL} have certain aspects in common. To start 
with, for all the cases we have checked, the rational parts of 
both equations are zero. It may be possible to prove this in 
general, but it is not obvious at all from expression 
\eqref{eq:wraphigherL}. For example, for $L=4$, the integral gives
\be
-\frac{640}{Q^5} + \frac{512}{Q^3} + \frac{4096Q(1+16Q^2 -16 Q^4)}{(1-4Q^2)^4} \ ,
\ee
where the first two terms obviously sum up to the result. The last 
term is responsable for the rational part, and sums up to zero. 
Going further, we have checked that up to at least $L=30$, the entire 
trancendentality structure of \eqref{eq:wraphigherL} coincides with what 
is expected for \eqref{eq:wrapZanon1/2}. That is, the $\zeta$ of minimal 
transcendentality is $\zeta (L-1)$ while that of maximal trascendentality 
is $\zeta (2L-3)$ \footnote{Such an $L$ dependence for the $\zeta$ of maximum transcendentality was conjectured in \cite{Suzuki}, for the $\mathcal{N}=4$ theory.}.

We can also extract the coefficient of the maximum trascendentality part 
and show that it coincides between the two expressions. According to the 
expansion displayed in \cite{Zanon}, up to at least $L=9$ the $\zeta$ of 
maximum transcendentality is always present in the coefficients $P_L$, $I_L^{(1)}$ 
and $I_L^{(2)}$, but not in the $I_L^{(j)}$ for $j>2$. Furthermore, in the exact all-loop 
expression for $I_L^{(1)}$, the highest 
trancendentality $\zeta$ enters as $\frac{1}{2}P_L$, and thereby cancels
the first $P_L $ coefficient in 
\eqref{eq:wrapZanon}. It therefore seems that the highest transcendentality 
contribution is determined by the 
$I_L^{(2)}$ coefficients. An all-loop form for these coefficients is also 
conjectured in \cite{Zanon}, with the result 
that the highest transcendentality $\zeta$ again has coefficient 
$\frac{1}{2}P_L$. All in all, inserting the 
definition of $P_L$, we have that the maximum trascendentality part 
of the first wrapping correction at $\beta = 1/2$ is
\be
\delta \gamma_{\text{ max. trans.}}^{1/2} = -64 g^{2L} 
\binom{2L-3}{L-1} \zeta (2L-3) \ . \label{eq:Emaxtrans}
\ee
In appendix \ref{app:maxtrans} it is proven that the maximal 
transcendentality contribution to \eqref{eq:wraphigherL} indeed 
coincides with this expression.\footnote{Since for large $L$, by Stirling's approximation,
\be
\binom{2L-3}{L-1} = \frac{L (L-1)}{2(2L-2)(2L-1)}\frac{(2L)!}{(L!)^2} \sim \frac{1}{8}4^L \ 
\ee
we see that \eqref{eq:Emaxtrans} will decay exponentially, at 
large $L$, as long as $g^2 < 1/4$. If none of the terms of less transcendentality 
grow faster than \eqref{eq:Emaxtrans} as $L \rightarrow \infty$, 
which is indeed the case for the values of $L$ that 
we have checked, the entire wrapping correction will decay exponentially. The large $L$ behaviour of the correction was later made rigourous and more precise in \cite{Beccariamagnon}.}

For odd $L$, the spectra at $\beta = 0$ and $\beta = 1/2$ are not expected to be related. Still, given the success for even $L$, we can try to extract some non-trivial information from the $SL(2)$-spectrum in this case as well. Unfortunately, this turns out to not be possible. In particular, for twist-three operators the spectrum is usually given in terms of harmonic sums evaluated at $M/2$ \cite{twist3} (the one-loop piece is proportional to $S_1\left(\frac{M}{2} \right)$), which do not seem to have meaningful values when analytically continued to odd values of $M$. As an alternative, one can use the relation\footnote{I would like to thank one of the referees of this article for pointing this out to me.} $S_1\left(\frac{M}{2}\right) = S_1(M) + S_{-1}(M)$, valid for even $M$, and continue the right hand side to $M=1$. The wrapping correction to twist-three operators, given in \cite{twist3wrap}, is as expected proportional to the square of the one-loop anomalous dimension and thereby vanishes when this prescription is used. Curiously, this coincides with the wrapping correction to the $\beta = 1/2$ magnon energy \eqref{eq:wrapZanon1/2} for odd $L$. However, the asymptotic expressions do not match since the one-loop energy at $\beta = 1/2$ is non-vanishing, so the odd $L$, $\beta = 1/2$ magnon does not appear in the $SL(2)$-spectrum. 

Instead, we can interpret the vanishing result at $M=1$ as giving a magnon of momentum $p=0$. Indeed, harmonic sums with positive indeces and argument $M/2$ can always be re-written as a combination of sums with argument $M$, having the property that the sums cancel at $M=1$. So if it continues to hold that the twist-three spectrum can be written in terms harmonic sums with positive indeces evaluated at $M/2$, then it should vanish when analytically continued to $M=1$ to all orders, in accordance with the energy of a $p=0$ momentum magnon. Furthermore, the Baxter equation at $M=1$ has no non-trivial solution for odd $L$, but it always has the trivial solution where all magnon rapidities are infinite, corresponding to zero momentum. In sum, the odd $L$ $SL(2)$-spectrum does not seem to provide any non-trivial information on the magnon energies as it, depending on the prescription used for analytical continuation, is either nonsensical or gives the trivial zero-momentum magnon at $M=1$.

Anyway, we see that there is a substantial amount of evidence that the first 
wrapping corrections to the magnon dispersion relation is the same for $\beta = 0$, 
$p=\pi$ and $\beta = 1/2$, $p=0$, for all even $L$, in agreement with the 
$\beta \rightarrow \beta +1/2$ symmetry. This gives a solid conjecture for the first wrapping correction to the single magnon operator \eqref{eq:op1magnon} at $\beta = 1/2$ for all $L$. For odd $L$ it is zero, and for even $L$ it is given by \eqref{eq:wraphigherL}.



We will conclude this section by using the $\beta \rightarrow \beta + 1/L$ symmetry 
to extract some information on the first wrapping correction to the 
magnon dispersion relation in $\mathcal{N}=4$ Yang-Mills. The symmetry implies 
that the anomalous dimensions of the single magnon operators, as given 
in \eqref{eq:wrapZanon}, at $\beta = N/L$, coincide with the energy 
of the $p=2\pi N/L$ magnon in the undeformed theory. For this set 
of momenta we thus have the wrapping correction
\ba
&& \delta E(p) = -2L g^{2L}\Big[ \big( C_{L,0}\left( \frac{p}{2\pi} \right) 
- C_{L,L-1}\left(\frac{p}{2\pi} \right) \big) P_L \nonumber \\
&& - 2\sum_{j=0}^{\left\lfloor \frac{L}{2} \right\rfloor - 1} 
\big(C_{L,j}\left(\frac{p}{2\pi} \right) 
- C_{L,L-j-1}\left(\frac{p}{2\pi} \right) \big) I_L^{j+1}  \Big] \ . \label{eq:dispwrap}
\ea
Curiously, we obtain the correction for precisely those magnons that correspond to physical (periodical) states of the spin chain.


\section{Conclusions}

\no
In this note we have given evidence that the symmetry 
$\beta \rightarrow \beta + 1/L$ of the $SU(2)_\beta$ spin chain of 
marginally $\beta$-deformed $\mathcal{N}=4$ Yang Mills, which 
is obvious from the point of view of the asymptotic Hamiltonian, continues 
to hold after wrapping corrections have been included. This was done by 
showing the equality of the physical magnon operator dimension at 
$\beta = 1/2$, and the $p=\pi$ magnon in the undeformed theory, for 
several spin chain lengths $L$, and for the term of maximum 
transcendentality to all $L$. It was also noted that previous gauge and string theory calculations are consistent with the symmetry.

The symmetry involves the general spin chain defined by relaxing the cyclicity 
constraint, but can still provide additional information on physical states. 
For example, equation \eqref{eq:wraphigherL} provides a compact expression 
for the first wrapping correction to the length $L$, with $L$ even, single 
magnon operator at $\beta = 1/2$. Furthermore, using the known expression 
for the wrapping correction to physical magnon operators at different 
$\beta$, the symmetry would determine the first wrapping correction to 
the $\mathcal{N}=4$ dispersion relation at momenta $p = 2\pi N/L$, for integer $N$.

The fact that we obtain the same magnon energy at $\beta = 0$ and $\beta = 1/2$ supports 
the assumptions made when calculating the energy 
of the $p=\pi$ magnon. It gives further weight, for example, that the energy of the 
$p=\pi$ magnon in the undeformed theory can be obtained, for even $L$, by analytically 
continuing the $SL(2)$ spectrum to $M=1$. Furthermore, the extension to arbitrary $L$ of the string thermodynamic Bethe ansatz inspired formula of \cite{Janik} gives the expected result.

One could object that the case that we have primarily checked, that is 
the coincidence of magnon energies at (shifted) momentum $\tilde{p}=\pi$ 
is special, since the string theory finite size factor $\cos \Phi$ 
becomes independent of the winding in that case. It could thus be 
that the symmetry does not hold in general, if for example the winding is 
chosen differently in gauge and string theories. However, as mentioned 
in section 2, the scaling dimension of the $L=4$ two-magnon operator, which is the Konishi operator in the undeformed theory, 
also satisfies the symmetry. And the momenta of the magnon consituents of the Konishi operator are $\pm 2\pi /3$ at lowest order, so this does not seem to be the case.

Curiously, these momenta are also special because up to 
integer multiples of $2\pi$, and apart from $p=0$, they are the only zeros 
of the wrapping correction \eqref{eq:dispwrap} to the dispersion relation. It would thus be valuable to have some additional checks of 
the symmetry. For example, the $L=6$, $M=3$ operators should have the same 
anomalous dimension at $\beta $ and $\beta + 1/3$.

It would also be interesting to try to connect the weak and strong coupling expressions of the magnon dispersion relations. At a first glance they seem to behave slightly differently. For example, the $\tilde{p}=\pi$, $\beta = 1/2$ magnon correction, given by \eqref{eq:wrapZanon1/2}, vanishes for all odd $L$, while the string theory correction of \cite{BykovFrolov} does not do so. This difference seems to carry over to the $\mathcal{N}=4$ theory, since the dimension of the magnon operator at $\beta = \frac{L-1}{2L}$ should coincide, by the symmetry, with the undeformed magnon energy at $p=\pi \frac{L-1}{L}$. Letting $L\rightarrow \infty$ this again implies that the first wrapping correction at weak coupling for odd $L$ at $p=\pi$ should vanish\footnote{Or at least have an additional power law suppression as compared to the even $L$ terms.} in an expansion around infinite volume. By contrast, the string theory finite-size result of \cite{N4finitesize} has no $L$-dependence beyond the argument of the exponential.


\vspace{5mm}
\centerline{\bf Acknowledgments}

We are grateful to L. Freyhult, for useful comments and discussions, and to C. G\'omez and especially R. Hern\'andez for much help and support during the various stages of the elaboration of this text. The work of J. G. is supported by a Spanish FPU grant.

\appendix

\section{The one-magnon wrapping correction}

\label{app:Janik}

\no
Here we will show how to obtain the expression \eqref{eq:wraphigherL}
for the first wrapping correction to the magnon of momentum $\pi$ in an 
$SU(2)$-sector spin chain of even length $L$. Our point of departure will be 
the formula $(17)$ of \cite{Janik}, giving the wrapping correction to the 
$M$-magnon state in the $SL(2)$-sector. We assume that, for the 
$\mathcal{N}=4$ theory, the magnon dispersion relation is global, 
in the sense that the first wrapping correction to the length $L$ 
$SU(2)$-magnon is the same as the first correction to the length 
$L-2$ $SL(2)$-magnon. By simply re-introducing the $L$ dependence of the leading part 
of the exponential factor
\be
\frac{4^{L-2}}{(q^2 + Q^2)^{L-2}} \ ,
\ee
where $L$ refers to the $SU(2)$ length, we have the wrapping correction 
\be
\delta \gamma(L,\, M) = -  4^{L-2} g^{2L} 
(\gamma_2^ L(M))^2\int_{-\infty}^\infty \frac{dq}{2\pi}\frac{T_M(q,\, Q)^2}{R_M(q,\, Q)}
\frac{1}{(q^2 + Q^2)^{L-2}} \ , \label{eq:wraphigherLgeneralM}
\ee
where 
\begin{align}
R_M (q,\, Q) = &P_M\left( \frac{1}{2}(q-i(Q-1)) \right)P_M\left( 
\frac{1}{2}(q+i(Q-1)) \right) \cdot \nonumber \\
&P_M\left( \frac{1}{2}(q+i(Q+1)) \right)P_M\left( \frac{1}{2}(q-i(Q+1)) \right) \ ,
\end{align}
\be
T_M(q,\, Q) = \sum _{j=0}^{Q-1} \left[ \frac{1}{2j-iq-Q} - (-1)^M 
\frac{1}{2(j+1) - iq - Q} \right]P_M\left( \frac{1}{2}(q-i(Q-1)) + ij \right) \ ,
\ee
$\gamma_2^L(M)$ is the coefficient of the one-loop energy, and where $P_M$ is the Baxter $Q$-function. The function $P_M (u)$ is proportional to
\be
\prod _{i=1}^M (u - u_i) \ ,
\ee
where the $u_i$ are the magnon rapidities solving the Bethe 
equations. For $L$ even, the one-loop Baxter equation requires 
the single magnon to have the root $u=0$, corresponding to 
momentum $\pi$, so in this case $\gamma _2^L(1)=8$, for all $L$, as seen from \eqref{eq:ek} or \eqref{eq:dispersion}. Furthermore, $P_1 (u)$ 
will be proportional to the identity. Requiring that 
\eqref{eq:wraphigherLgeneralM} at $M=1$ gives the wrapping 
correction \eqref{eq:janik4loopm1} when $L=4$, 
fixes the constant of proportionality to $2i$. With the 
replacement of the Baxter function by the identity, things 
simplify considerably and we have
\be
R_1(q,\, Q) = (q^2 + (Q-1)^2)(q^2 + (Q+1)^2) \ , \label{eq:R1}
\ee
and
\be
T_1(q,Q) 
= \! - \! \sum _{j=0}^{Q-1} \! \left[ 2 + \frac{i}{q-iQ+2ij} - \frac{i}{q - iQ + 2i(j+1) } 
\right] \!
= \! -2 Q \! \left( \! 1- \frac{1}{q^2+Q^2} \! \right)  . \label{eq:T1}
\ee
Introducing \eqref{eq:R1} and \eqref{eq:T1} 
into \eqref{eq:wraphigherLgeneralM} for $M=1$, and using that, 
according to \cite{Janik}, the only pole that contributes to the 
integral after summing over $Q$ is the one at $q=iQ$, gives the desired result.


\section{The maximal transcendentality contribution to the first wrapping correction}

\label{app:maxtrans}

\no
In this appendix we will show that the maximal transcendentality term of 
\eqref{eq:wraphigherL}
coincides with the maximal transcendentality piece in the marginally deformed 
theory, \eqref{eq:Emaxtrans}.
Let us calculate the relevant part of the residue by noting  
that the term of maximal transcendentality 
dominates in the limit $Q \rightarrow 0$. When expanding in terms 
of $(q-iQ)$, the expansions of the $1/(q^2 + (Q \pm 1)^2)$ factors 
will be subleading in this limit, and can simply be set to one 
(which is what they evaluate to at $q=iQ$, when $Q\rightarrow 0$). For example,
\be
\frac{1}{q+i(Q-1)} = \frac{1}{i(2Q-1)}\sum _{j=0}^{\infty} i^j \left( \frac{q-iQ}{(2Q-1)} \right)^j \ ,
\ee
for which the expansion coefficients do not receive an enhancement 
as $Q \rightarrow 0$, while the factor $1/(q+iQ)$, present in $1/(q^2 + Q^2)$, has the expansion
\be
\frac{1}{q+iQ} = \frac{1}{2iQ}\sum _{j=0}^\infty i^j \left( \frac{q-iQ}{2Q} \right)^j \ .
\ee
The most singular term is then the one having the maximum number 
of $1/(q-iQ)$ terms, since these have to be compensated for by the 
enhanced $1/(q+iQ)$-terms. We can thus substitute the numerator 
$\left(1 - 1/(q^2 + Q^2)\right)^2$ for $1/(q^2 + Q^2)^2$. So the maximal 
transcendentality contribution is simply obtained by evaluating the residue
\be
\text{Res}_{q=iQ}\left[ \frac{Q^2 }{ (q^2 + Q^2)^{L} } \right] \ ,
\ee
which is equivalent to finding the $(L-1)-th$ expansion coefficient 
of $Q^2/(q+iQ)^L$. This, in turn is given by
\be
\frac{Q^2}{(2iQ)^{2L-1}}\binom{-L}{L-1} = \frac{1}{4^{L-1}}
\frac{1}{i}\frac{1}{Q^{2L-3}}\binom{2L-3}{L-1} \ . \label{eq:maxtransres}
\ee
Substituting the residue of \eqref{eq:wraphigherL}
for \eqref{eq:maxtransres} 
and summing over $Q$ immediately gives \eqref{eq:Emaxtrans}.


\end{document}